# Investigating introductory and advanced students' difficulties with change in internal energy, work and heat transfer using a validated instrument

Mary Jane Brundage[1], David E. Meltzer [2], and Chandralekha Singh[1]
[1] *Department of Physics and Astronomy, University of Pittsburgh, Pittsburgh, PA, 15260*
[2] *College of Integrative Sciences and Arts, Arizona State University, Mesa, AZ 85212*

We use the Survey of Thermodynamic Processes and First and Second Laws-Long (STPFaSL-Long), a research-based survey instrument with 78 items at the level of introductory physics, to investigate introductory and advanced students' difficulties with internal energy, work, and heat transfer. We present analysis of data from 12 different introductory and advanced physics classes at four different higher education public institutions in the US in which the survey was administered in-person to more than 1000 students. We find that not only introductory but also advanced physics students have many common difficulties with these introductory thermodynamic concepts after traditional lecture-based instruction in relevant concepts. We utilize a wide variety of problem types and contexts and our sample includes large numbers of introductory algebra-based, calculus-based, and advanced students. Some of our findings are consistent with prior research in this area, but others expand upon them and reveal previously unreported aspects of students' thinking. Findings related to common difficulties of students before and after traditional lecture-based instruction in college physics courses can help instructors of these courses plan instruction and curricula to improve student understanding. These findings can also be valuable for developing effective research-based curricula and pedagogies to address student difficulties and help students develop a functional understanding of these fundamental thermodynamic concepts.

# I. INTRODUCTION AND GOAL OF THE INVESTIGATION

Many physics courses for science and engineering majors focus on helping students with both conceptual understanding and problem solving. In this context, research-based conceptual multiple-choice surveys can be invaluable for investigating student understanding of physics before and after instruction using various curricula and pedagogies [1-20]. These surveys allow administration of large sets of concept-focused, non-quantitative questions to large numbers of students simultaneously, and—especially when linked to one-on-one problem-solving interviews with individual students—can provide information regarding students' ideas and reasoning processes that can be utilized in planning instructional materials and methods. By identifying response patterns to the question sets it is often possible to gain insight into the types of problems that students find difficult, why they find them to be difficult, and how responses on different types of problems are related to each other. These surveys—so long as they are combined with individual student interviews—have the potential to yield a broader and deeper picture of students' reasoning on specific physics topics than is possible with just one or a small handful of diagnostic questions.

The topic of thermodynamics is a core component of many introductory physics curricula, and prior research suggests that both introductory and upper-level students have many difficulties with introductory thermodynamics concepts, thus justifying the need for a diagnostic instrument to monitor the progress of student learning on these topics [15-51]. Earlier we investigated students' thinking using the STPFASL-Short survey with 33 items [49-51] in which some of the concepts (e.g., change in internal energy, work, and heat transfer) are combined into a single question, making it difficult to disentangle students' thinking with each thermodynamic variable separately. Here, by contrast, we discuss use of a research-based, validated 78-item multiple-choice survey instrument called the Survey of Thermodynamic Processes and First and Second Laws-Long (STPFaSL-Long); in this instrument, each item focuses on one specific concept (e.g., change in internal energy in isothermal processes), greatly increasing its power to reveal specific student ideas. We administered this instrument in traditional lecture-based introductory and upper-level undergraduate physics courses and to physics graduate students in their first-year, first-semester, to investigate the extent to which students could answer questions on basic concepts posed in different contexts. The survey explores students' thinking regarding changes in internal energy, work done by and on a system, and heat transfer to and from a system, at a level typical of thermodynamics instruction in introductory college physics courses, with survey data collected from 12 different courses at four different universities in the US. All upper-level students were once introductory students, so data regarding their difficulties with conceptual problems can potentially be helpful in meeting the needs of students in both introductory and advanced physics courses. The details pertaining to the development, validation, and administration of the STPFASL-Long survey can be found elsewhere [52]. In addition to administering the written survey in various courses, we interviewed 11 introductory and 6 upper-level students individually using a think-aloud protocol to get a deeper insight into students' thought processes as they answered the survey questions.

The 78-item survey we developed and administered is not only the largest research-based conceptual thermodynamics survey ever created specifically for use in physics courses, but also one of the largest physics concept surveys on any topic. The breadth of coverage and the large variety of question contexts together offer unparalleled power in mapping out students' thinking on key thermodynamics concepts that are at the core of introductory physics courses. Our findings, presented in detail below, serve to confirm and validate a number of results previously reported in the physics education literature, as well as to provide numerous fresh insights into students' thinking that have never been previously reported, or which have only been hinted at in previous reports.

Regarding the framework, in our discussion, we will often employ the term student "difficulty," a term most often associated with and frequently used by the University of Washington Physics Education Group. In our analysis, the meaning we attach to this term is essentially the same as theirs, which is *use of a specific idea or pattern of reasoning instead of those we consider correct and appropriate* [53]. More broadly, we are trying to find out "what ideas students have before [or after] traditional lecture-based instruction that may make it challenging for students to develop a sound conceptual understanding" and "what ideas can be built upon to promote student learning, e.g., using research based curricula and pedagogies." The way we carry out this investigation is to pose to students numerous conceptual question set in diverse contexts that all relate to basic principles of thermodynamics as typically taught in introductory physics courses. By studying both the prevalence of correct and incorrect responses, and the *specific patterns* of such responses, we develop significant insight into students' thinking processes. Instructors and curriculum developers can then make use of this information to inform their work with the goal of improving the effectiveness of instruction.

A number of the conceptual difficulties in the research discussed here have previously been documented [15-51], at least to some extent, through use of a limited number of problem types and contexts. The great diversity of problem contexts in our survey has allowed us to differentiate the *specific* problem types and contexts found most challenging by students, and to significantly sharpen the precision of our understanding of students' thinking. In some cases, we demonstrate the robustness of previous findings, and in others we reveal nuances that were previously unexplored. In a large number of instances, we report new findings that are absent from the extant literature. Our large sample size of over 1000 students from four different universities provides unprecedented statistical power in quantitatively determining the prevalence of specific response patterns. Our inclusion of both calculus-based *and* algebra-based introductory students, as well as a large sample of upper-level students, has revealed thinking patterns and comparisons between algebra-based and calculus-based students, and between introductory and upper-level students, that were previously unknown due to the relative rarity and small sample sizes of comparable studies. The tables presented in this paper provide detailed quantitative data pertaining to students' survey responses for all three student groups, while our narrative combines interview data and survey responses to explore introductory and upper-level students' thinking in greater depth and detail.

A number of prior studies [15-51] have focused on student understanding of thermodynamics; here we provide a few examples of investigations related to students' ideas about internal energy, work, and heat transfer. Loverude et al. [21] investigated student understanding of work in the context of adiabatic compression of an ideal gas. In one problem used in their study, students were asked to consider a cylindrical pump containing an ideal gas such that the piston fit tightly, and no gas could escape. In one version, students were asked what would happen to the gas temperature if the piston were quickly pressed inward; students were also asked to explain their reasoning. Another type of problem posed in the same study related to a cyclic process represented on a PV diagram in which various parts of the cyclic process were isothermal, isochoric, and isobaric. Students were asked whether the work done in the entire cycle was positive, negative, or zero, and to explain their reasoning; about half said that the net work was zero. Loverude et al. [21] reported that students had difficulty in discriminating between related concepts, sometimes considered only two of the three variables in the first law of thermodynamics, and sometimes confused heat and internal energy or were confused about how work, heat, and internal energy were related. A different study of students' understanding of heat, work, and the first law in an introductory calculus-based physics course by Meltzer [23] explored students' thinking on several conceptual problems, some of which involved PV diagrams. For example, one problem in that study involved two different processes represented on a PV diagram that started at the same point, *i*, and ended at the same point, *f*. Students were asked to compare the work done by the gas and the heat absorbed by the gas in the two processes and to explain the reasoning for their answers; very low correct-response rates were found. Meltzer also reported that students often incorrectly claimed that work $W = 0$ and thermal energy transfer $Q = 0$ for a cyclic process, and had difficulty distinguishing between state variables and process-dependent variables [23]. Leinonen et al. focused their investigations on the pre-knowledge of introductory thermal physics at the university level [26], on student understanding of adiabatic compression of an ideal gas [27], and on how hints and peer instruction during lecture might help students [28]. Most of their findings were consistent with Loverude et al. [21] and Meltzer [23] in that evidence-based instructional approaches can help improve student understanding. Meltzer also investigated upper-level students' understanding of these concepts using similar measures and found that they also had similar difficulties (e.g., belief that $W = 0$ and $Q = 0$ in a cyclic process), but that the prevalence of these types of difficulties were reduced in comparison to introductory students [25]. In particular, upper-level students demonstrated qualitative reasoning skills superior to those of introductory students and were better at interpreting the meaning of diagrams and graphs [25].

Below, we first describe the methodology for our research followed by presentation of results and discussion. We first present data regarding students' survey response on items related to internal energy $E$ as a state variable vs. $W$ and $Q$ as path-dependent variables, followed by other data reflecting students' responses on other diagnostic items related to $E$, $W$ and $Q$. Finally, we conclude with a summary of the findings and instructional implications.

## II. METHODOLOGY

The Survey of Thermodynamic Processes and First and Second Laws-Long (STPFaSL-Long), a validated survey instrument with 78 items, was used in this research. This instrument focuses on introductory thermodynamics concepts. The details of the development and validation of the STPFaSL-Long survey instrument can be found in Ref. [52] and the survey can be found in Ref. [54]. Most items (conceptual questions) on the survey have four possible answer choices and are related to one of several different thermodynamic variables, including among others internal energy, work, and heat transfer. The answer options typically ask whether the thermodynamic variable is positive, negative, or zero (or increases, decreases, or remains the same, as appropriate) during a specified thermodynamic process, or whether there is not enough information provided to determine the answer in the given situation. There are 22 out of 78 items that are true/false (T/F) questions.

This investigation uses survey data obtained both before and after traditional lecture-based instruction (pre-test and post-test, respectively) in relevant concepts. In particular, the written data analyzed here were taken by administering the survey in proctored in-person classes as a pre-test (before instruction) and post-test (after students had learned the relevant concepts), but before students' final exam in the course. Students were given some extra credit for completing the survey. These written student data are from 12 different in-person courses from four different large public institutions; students completed the survey in class on Scantrons during a 50-minute class period if they took the entire survey since some students were administered only the first 48 or last 52 questions [52]. All classes primarily employed traditional lecture-based instruction. We discuss analysis of student difficulties based on the written data from five groups of students: (1) 550 students in the introductory algebra-based (Int-alg) physics course after instruction (post-test); (2) 371 students in the introductory algebra-based (Int-alg) physics course before instruction (pre-test); (3) 492 students in the introductory calculus-based (Int-calc) physics course after instruction (post-test); (4) 753 students in the introductory calculus-based (Int-calc) physics course before instruction (pre-test); and (5) 89 students in their upper-level thermodynamics courses after instruction (post-test). However, not all introductory students answered all survey questions, and some introductory recitations were given only the first 48 or last 52 questions to ensure split-test reliability [52]. On the post-test (pre-test), out of the 492 (753) Int-calc students, 168 (505) were given the first 48 questions, 73 (248) were given the last 52 questions, and 251 (0) were given the full survey. On the post-test (pre-test), out of the 550 (371) Int-alg students, 170 (173) were given the first 48 questions, 218 (198) were given the last 52 questions, and 162 (0) were given the full survey. The average scores of all the in-person introductory student groups were in the range 51-58%; the upper-level group who scored 76%. The standard deviation in students' individual scores for all introductory groups ranged from 9%-13% while the standard deviation in scores of the upper-level students was 14%; thus, standard errors for scores on the full instrument were on the order of 1%. Students in the Int-calc courses were typically engineering majors with some physics, chemistry, and math majors, while students in the Int-alg courses were mainly biological science majors and/or those interested in health-related professions. Students included in the upper-level group were typically physics majors in thermodynamics courses or Ph.D. students in their first year, first-semester of their graduate program, who had not taken any graduate-level thermodynamics. (Since the survey was administered as a pre-test to these upper-level students, they were presumed to have taken upper-level undergraduate thermodynamics). More information on the survey can be found in the paper discussing the validation of the STPFaSL-Long [52].

The interview data are from 11 introductory and 6 upper-level students from one institution who volunteered after an opportunity to participate in this study was announced. Each interview lasted between 1-2 hours in one sitting depending upon students' pace. The interviewed students were given $25 for their participation. The interviews used a semi-structured think-aloud protocol. Students were asked to think-aloud as they answered the survey questions and were not disturbed except for being urged to keep talking if they became quiet. Only at the end did we ask them for clarifications of points they had not made clear, particularly if they did not answer questions correctly. Lastly, 349 students from two Int-calc courses were asked to answer the survey questions at the beginning of the semester (pre-test) electronically on Qualtrics and provide their reasoning for each question. While many students did not provide meaningful reasoning, some students provided very short but insightful

responses. We will only discuss these written explanations for various survey items for cases in which most interviewed students provided the correct responses, such that the interviews (in those cases) did not provide sufficient insight into reasons for student difficulties.

## III. RESULTS AND DISCUSSION

Table I (below) provides the breakdown of response rates on various survey items; correct-response rates are indicated in boldface. Other tables referenced in the text may be found in the Appendix.

A. Difficulties with state variables
   1. Difficulties with state variables in the context of cyclic processes

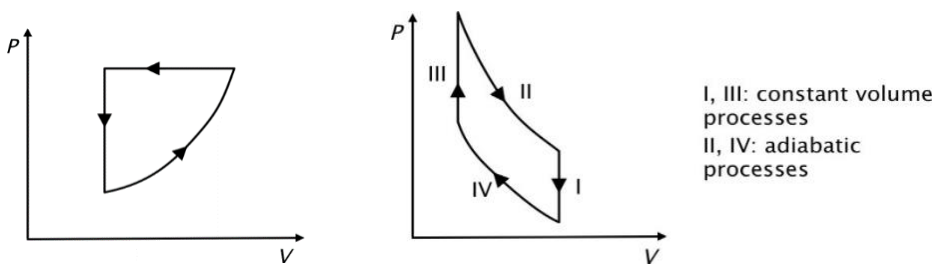

**FIGURE 1.** The diagrams for items 6-9 (left) and items 24-26 (right) on the survey.

Several prior investigations have studied student understanding of thermodynamic concepts in cyclic processes that start and end in the same state [22, 23, 30, 47]. Fig. 1 includes diagrams for survey items 6-9 (left) and items 24-26 (right) showing counterclockwise and clockwise cyclic processes, respectively. Since internal energy $E$ is a state variable, it is unchanged over the course of one complete cycle (i.e., $\Delta E = 0$). On the other hand, the work done by the system $W$ and the heat transfer to the system $Q$ are path dependent and will differ depending upon the actual path traversed, even if the initial and final states of the system are the same. Our findings align with those of earlier studies. In particular, consistent with prior studies, we find that many students believed that $W$ and $Q$ would be zero for the cyclic processes, thus failing to distinguish their properties from those of state variables.

TABLE I. Correct-response rates as percentage of total responses (in boldface) and rates of various incorrect responses for students in upper-level, introductory calculus-based (Int-calc), and algebra-based (Int-alg) physics courses for problems related to internal energy, work, and heat transfer in cyclic processes and processes that share the same initial point $i$ and same final point $f$. Survey item numbers and answer options consistent either with the correct response or with the specified student difficulty are provided in each row. Item #s with an asterisk (*) are T/F questions.

| **Correct answers in bold**, incorrect answers unbolded | Item # | Answer Choices | Prevalence (%) by level | | | | |
|---|---|---|---|---|---|---|---|
| | | | Upper Post | Int-calc Post | Int-calc Pre | Int-alg Post | Int-alg Pre |
| **Treating the internal energy $E$ as a state variable (correct)** | 6 | **A** | **89** | **80** | **79** | **80** | **80** |
| | 25 | **C** | **83** | **72** | **63** | **69** | **72** |
| | 47* | **A** | **84** | **71** | **72** | **72** | **73** |
| Not treating the internal energy $E$ as a state variable | 6 | B+C+D | 11 | 20 | 21 | 20 | 20 |
| | 25 | A+B+D | 17 | 28 | 38 | 31 | 28 |
| | 47* | B | 16 | 29 | 28 | 28 | 27 |
| **Correct response on work $W$** | 7 | **B** | **51** | **46** | **13** | **18** | **5** |
| | 46* | **B** | **90** | **74** | **63** | **68** | **65** |
| | 7 | C | 16 | 21 | 59 | 51 | 71 |

| | | | | | | |
|---|---|---|---|---|---|---|
| Treating the work *W* as though it is a state variable | 46* | A | 10 | 26 | 37 | 32 | 35 |
| **Correct response on heat transfer** ***Q*** | 9 | C | 44 | 35 | 21 | 20 | 16 |
| | 12 | C | 47 | 29 | 23 | 23 | 17 |
| | 26 | B | 52 | 31 | 16 | 16 | 18 |
| | 30* | B | 85 | 67 | 55 | 55 | 50 |
| Treating the net heat transfer *Q* as though it is a state variable | 9 | A | 16 | 33 | 44 | 56 | 58 |
| | 12 | A | 28 | 56 | 67 | 66 | 72 |
| | 26 | A | 25 | 37 | 49 | 55 | 61 |
| | 30* | A | 15 | 33 | 45 | 45 | 50 |

Table I summarizes student responses to questions dealing with the concept of a state variable. Items 6, 7 and 9 on the survey (Fig. 1 left) relate to *E, W*, and *Q*, respectively, in a counterclockwise cyclic process, and items 25-26 (Fig. 1 right) relate to *E* and *Q*, respectively, in a "reversible [clockwise] cyclic process." The results show that while most students (63% or more) gave correct responses both pre- and post-instruction on the questions related to internal energy *E*, only the upper-level students were able to reach 50% or more correct responses on any of the questions related to heat *Q* and work *W*.

The lowest correct-response rates were for the introductory student groups on the heat-related questions (35% or fewer correct responses, pre- or post-instruction), again consistent with findings from previous studies. Students in the algebra-based courses performed significantly worse than those in the calculus-based courses on questions related to *Q* and *W*. More than a third of the calculus-based students (33-37%), and more than half of the algebra-based students (55-56%), asserted *after instruction* that net heat transfer during the cyclic process would be zero. It is interesting that on a related question (item 30) that simply asks whether it is true that "there is NO net heat transfer between the system and the surroundings" in a cyclic process, correct-response rates in all student groups were 50% or greater. However, on this true/false question, rates close to 50% suggested that most introductory students were merely guessing at the correct answer. On item 7, the single cyclic-process question related to work, more than half (51%) of the Int-alg students said that the net work done by the gas for one complete cycle would be zero, a response given by 21% of the Int-calc students. A sign error was the most common error among the Int-calc students. (See further discussion below in Section C2.) These results are consistent with those reported in Refs. [21, 23] and broaden them by demonstrating consistency across a wider range of problem contexts.

The interview data shed additional light on students' thinking related to the survey items. It is striking that, in many of their responses, students invoked relationships between only two of the three variables (*Q, E,* and *W*) that are incorporated in the first law of thermodynamics, leading them to make incorrect inferences. This "variable-exclusion" thinking pattern related to thermal phenomena was previously described by Rozier and Viennot in 1991 [55], as well as in Ref. [21]. For example, one student, responding to the cyclic-process question in item 25, stated "…because the work is positive, it [final *E*] would be less than the initial internal energy," thus evidently ignoring the role of heat transfer in the process. In interviews regarding heat-related questions, the commonality of initial and final states was often invoked to assert that net heat transfer would be zero in the cyclic process. Responding to item 9, one student said, "In that case, heat transfer…there is none since we are coming back to the original condition at the end of the cycle." In response to item 26 related to *Q* in a clockwise cyclic process (Fig. 1 right), an introductory student said, "But if you're returning to the same state, so would there be no change 'cause you are finishing where you started? I feel like that sounds more right." In response to item 30, which is a true/false question, an introductory student stated, "Any cyclic process, I think this is true [*Q* = 0] since there is going to be no net heat transfer, there is no net heat added." Responding to the same question, another introductory student said, "Any cyclic process, there probably is, oh, no net heat transfer. Sure. I guess cyclic, you end up in the same place…" In response to item 46, which was a true/false question about whether *W* "is determined by the state of the system and not by the process that led to the state," one student stated, "…I think for work, it should be yes because we can measure the pressure and volume [in a state on which *W* depends]."

In Ref. [23], interview responses regarding heat and work done in a cyclic process reflected thinking analogous to that reported above, i.e., that the identical values of temperature, pressure, and volume in initial and final states implied that both net heat transferred, and net work done would be zero. While the most popular incorrect response for introductory students was that net heat transfer would be zero, a substantial number of students did in fact recognize that net heat transfer would be non-zero but made a sign error in their response. Interviews suggested that these sign errors arose either from confusion about whether net work was done *on* or *by* the system, or by reasoning errors in connecting the sign of the work done to the sign of the net heat transfer. To our knowledge, this finding that the "work done *on* or *by*" confusion extends to the context of PV diagrams has apparently not been previously reported.

2. Difficulties with state variables when two processes start and end at the same points

Table I also shows student responses on item 12 (Fig. 2) in which an ideal gas undergoes two different processes, both of which start in the same initial state, $i$, and end in the same final state, $f$; the processes are shown to have different paths on the PV diagram. Students are asked to compare heat transfer $Q$ for the two processes, that is, to state whether $Q$ for Process 1 is greater than, equal to, or less than $Q$ for Process 2. On this challenging item, drawn directly from Ref. [23], well over half of the introductory students (and over a quarter of the upper-level students) responded incorrectly that heat transfer $Q$ is equal in Processes 1 and 2. Correct-response rates were less than 30% for introductory students and only 47% for upper-level students, comparable to the rates on the cyclic-process items (9 and 26) that also dealt with heat transfer.

Correct-response rates and interview responses were consistent with those reported in Ref. [23]. For example, one of the introductory students in the present study stated, "I think that heat transfer should be equal because they have the same starting and ending point. And then it's just the difference in pressure and volume." Another student said, "I'm thinking if I can use that $E=Q-W$ equation again…I'm just going to say that it's equal [$Q$ for both processes] because they start and end in the same place. Yeah."

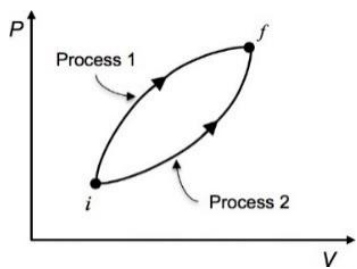

**FIGURE 2.** The diagram for item 12 on the survey

3. Difficulties with state variables in general

Items 46 and 47 are yes/no questions that simply ask whether work done by the system (item 46) and internal energy of the system (item 47) "are determined by the state of the system and not by the process that led to that state." Correct response rates on these items for the introductory students were better than random guessing but not by very much. Only among the upper-level students did correct-response rates on items 46 and 47 (90% and 84%, respectively) indicate convincingly that most students knew the correct answer. Interview responses supported the previous observation that "variable-exclusion reasoning" was at work, for example: "Wait, but that depends on work, so no again [$E$ is not a state variable]"; "…internal energy…I'm going to also say no [$E$ is not a state variable] because work is in the internal energy equation." Both these interviewees ignored the role of heat transfer.

B. Other difficulties with $\Delta E$

Below we describe difficulties other than those associated with $E$ being a state variable, which was discussed in the preceding section III.A:

1. Not recognizing that $E$ is proportional to the absolute temperature, $T$, for an ideal gas

There are three items whose correct answer depends only on one single concept, which is: For an ideal monatomic gas, the internal energy of the system, *E,* is proportional to the absolute temperature, *T*. We note that the mathematical relationship between *E* and *T* of an ideal gas is provided explicitly on the cover page of the survey for students' reference: $E_{int} = (3/2)NkT$. It is remarkable that correct-response rates on the three items (shown in the Appendix, Table II) vary so significantly, implying that students' grasp on this concept is substantially weaker than might be inferred from results on only a single test item. In items 33, 34, and 69, temperatures of the initial and final states of two processes are *explicitly* provided on the *PV* diagrams; in 33 and 34, students are asked whether the internal energy increases, decreases, or does not change during each process (see Figs. 3 and 4). For the process in which temperature increases (item 33), a large majority of students (76%/81% for calc/alg) give a correct post-instruction response, while only about half (58%/55%) do so on the process in which temperature *decreases* (item 34). There was actually little difference between the pre- and post-instruction scores on these items. But on item 69, in which students are asked to compare the amount of internal energy change in two processes (one adiabatic and one isochoric) that share a common initial state and equal-temperature final states, less than a third of the students (32%/25%) answer correctly that the internal energy changes are equal. From this we conclude that most introductory students' understanding of the proportionality of *E* and *T* for an ideal gas is weak at best, and that contextual features of a problem can thus easily derail the problem-solving process. Although the difficulty with the *E-T* proportionality has been previously reported, the strong context dependence of the response rates—even when temperatures are *explicitly* provided—is a new finding. The same response-rate pattern is reflected among the upper-level students as well, although a clear majority (60%) of that group can give a correct response even on the most "difficult" item, #69.

Although our results are consistent with those of earlier studies, e.g., Refs. [22, 26, 27, 30, 39, 50, 51], responses given during our student interviews provide additional insight into students' thinking and the roots of their evident confusion on this concept. It is clear that many students were grasping for clues from properties of the specific processes, but largely ignoring the definitive evidence provided by the single relevant parameter, the temperature *T*. For example, an introductory student stated regarding item 69, "Ok, ok they're not equal because process 2 does not have work so because of that I'm going to say that for process 2 [isochoric], the change in the internal energy is less than process 1 [adiabatic]. So greater in process 1 than in process 2." An upper-level student who started out on the right track became confused, concluding that there was not enough information: "So process 1 is adiabatic which means *Q* is equal to zero and there is some positive work. Process 2 is isochoric which means work is zero and *Q* is something…Change in internal energy for the first one is just the work done for the first one. And then for the second one, change in internal energy is just the heat from the second one. Change in internal energy of the gas is equal? Not necessarily. It depends on what the value of work is for this and the value of heat is for this…but both the temperatures are actually increasing. I'll just say not enough information, since I don't know what the values of those two terms are."

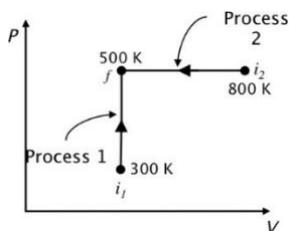

**FIGURE 3.** Diagram provided for items 33 and 34.

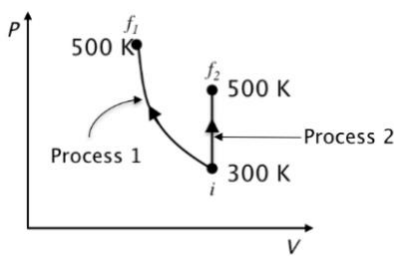
**FIGURE 4.** Diagram provided for items 69 and 70.

2. Not recognizing that $\Delta E = 0$ for an isolated system

The internal energy $E$ of a system cannot change in any process in which there is neither work done by or on the system, nor heat transferred to or from it. The challenge for the student is in recognizing when those conditions hold. There are four survey items (13, 23, 50, and 72) involving processes in systems consisting of two interacting objects or gas containers, and students are asked whether the internal energy $E$ of the system changes during the process; previous investigations have not explicitly analyzed students' thinking about $E$ in this important context. In all four cases, the systems are thermally isolated and zero work is done, so $\Delta E = 0$. Table II (Appendix) shows that most students in all three groups gave correct responses both pre- and post-instruction. The highest correct-response rates were on item 13 involving thermal interactions between two solid blocks, for which it is particularly obvious that net work is zero. By contrast, interviews related to item 23 (the free expansion of a gas into an evacuated container) suggested that incorrect responses were often due to the incorrect idea that work is done during a free expansion. (Further discussion of this specific difficulty can be found below in Section C.1.) For example, one introductory student said "I think the internal energy of the system should decrease. Well, not actually sure why, but I just get the feeling that would happen because no heat is added. I don't think there is any heat transfer between the surroundings and the system, but the gas does work. Right, so that should mean internal energy should decrease." A variant of this thinking was expressed by an upper-level student, who thought "internal energy of the system will decrease because the volume is now twice the volume that it was before," apparently associating this volume increase with quasistatic expansion processes in which net work *is* done by the system. Interview responses on the other items suggested confused attempts to simultaneously compare changes in multiple thermodynamic variables, in each case overlooking the key fact that there was neither any net heat transfer nor any work done during these processes.

3. Not recognizing that $\Delta E = 0$ for an ideal gas undergoing an isothermal process

The internal energy of a monatomic ideal gas is proportional to temperature so it will not change during an isothermal process. However, as discussed above in Section B.1, most students appear to be either unfamiliar with this idea or unable to apply it consistently to actual processes. Results shown for items 61 and 65 in Table III in the Appendix are consistent with this conclusion, as less than half of all introductory students agreed post-instruction that internal energy would not change in processes explicitly identified as isothermal (in 61, an "isothermal expansion," in 65, a "reversible isothermal expansion"). Even upper-level students had difficulty with these items, as more than one third of them gave incorrect responses.

The detailed results on these two items allow us to drill down somewhat on the specific difficulties involved. One might have thought that the internal energy loss due to work done in an expansion process would lead students to conclude that internal energy would decrease (by ignoring heat transfer), and indeed interviews showed that some students did follow this line of reasoning. For example, one introductory student said, "Internal energy of the gas during a reversible expansion…Um…it will decrease since there is some work being done, positive work by the gas done. So then the internal energy will decrease." An upper-level student argued, "So $U=Q-W$, is there work done? It expands, so I guess there is work done. So we set it to $Q=0$ so the internal energy of the gas has to decrease because the expansion with it is getting larger, so $\Delta V$ is positive." One interviewed student seemed to confuse isothermal and adiabatic processes: "So because the change in entropy would be zero, I'm going to say that the internal energy decreases because there would not be a heat transfer and that would just leave a negative work." In fact, interviews reported in Ref. [23] showed that many students are indeed unaware that heat transfer does take place during an isothermal process. Other investigators had previously reported analogous findings. Our study indicates that even decades later, similar difficulties persist among introductory physics students after traditional lecture-based instruction and furthermore, that this difficulty often leads students to conclude that internal energy would *decrease* during an isothermal expansion.

Despite the evident attractiveness of the "expansion means $E$ decreases" argument, the more popular incorrect response among all groups was that internal energy would *increase*, rather than decrease. Interview responses reflected a variety of thinking processes that led to this incorrect conclusion. For example, one student stated, "Internal energy…if the gas is expanding, then I think volume would increase, so I think the internal energy increases." Another argued, "So if it's expanding and that means work would be positive, and I said it got heat into the gas, so that heat is also positive. So I think internal energy would also increase because of that." A handful of students (5% or less) concluded that there was not enough information to decide the question asked. It seems clear that many students were led into circuitous and unfruitful lines of reasoning because they simply did not connect the ideas that "isothermal" means "no temperature change," and that constant temperature means unchanging internal energy for a monatomic ideal gas. If students do not have those ideas clearly in their knowledge structure, it is possible for them to go on a variety of extended chains of unproductive reasoning.

4. Incorrectly thinking that *ΔE* = 0 or *ΔE* > 0 for an adiabatic expansion

In an adiabatic expansion, since heat transfer is zero, internal energy will decrease due to work being done by the gas. Results on item 2 in Table III in the Appendix show that fewer than one third of introductory students—and barely more than half of upper-level students—correctly responded that internal energy must decrease in a "reversible adiabatic expansion." The most common response among introductory students (Int-calc: 39%; Int-alg: 47%) was that internal energy remains constant. Slightly more than a quarter of introductory students made a sign error, concluding that internal energy would increase rather than decrease. Interviews indicated that students often completely ignored work done by the gas as a factor in internal energy calculations. For example, one student argued, "It's reversible, ok, internal energy…it's expanding, I feel like the internal energy would increase. Yeah." Another student stated, "I believe there…um…the internal energy remains constant because there's no heat exchange…in the adiabatic expansion process." Yet another student argued that $\Delta E=0$ "because [there's] no heat going between the system and the surroundings." These are examples of student reasoning during interviews that simply ignored work.

Loverude, Kautz, and Heron reported the great difficulties that students encountered when attempting to apply the work concept in thermodynamic contexts [21]. They showed that few students were spontaneously able to invoke the concept of work when discussing the adiabatic compression of an ideal gas. In fact, students were unable to understand that an entity called "work" could bring about a change in the internal energy of a system. Further confirmation of these observations was reported in Ref. [23]. Our present study reconfirms, greatly amplifies, and expands upon those previous findings.

5. Incorrectly thinking that *ΔE* = 0 or having incorrect sign of *ΔE* for an isobaric process

We have already discussed results on item 34 in Section B.1 above, in which students are asked about internal energy changes in an isobaric process in which both temperature and volume decrease. We argued that the high error rate on this item (over 40% for introductory students), in the context of similar problems in which temperatures were explicitly shown on PV diagrams, was reflective of students' weak grasp of the proportionality relationship between temperature and internal energy. Item 45 shows another isobaric process (Fig. 5), this one involving a volume increase; however, the initial and final temperatures are not shown on the PV diagram in this case, in contrast to item 34. Correct-response rates on item 45 (shown in Table III in the Appendix, along with those of item 34) were substantially lower than on item 34, with well under half of introductory students able to provide a correct answer. Sign errors were more common than on item 34, as were responses that internal energy would not change at all. On both items, sign errors were the most common incorrect response.

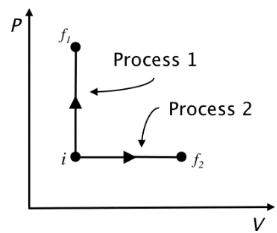

**FIGURE 5.** Diagram provided with problems 40-45

Interview responses suggest that analysis of $\Delta E$ for the isobaric process is challenging for students partly because all three thermodynamic variables in the first law are non-zero. Students sometimes did "two-at-a-time" or "variable exclusion" reasoning when applying the first law of thermodynamics [50, 51]. In response to item 34, some students did not consider the heat transfer, even though all three thermodynamic variables are non-zero in an isobaric process. As we have repeatedly emphasized throughout this section, students' difficulties on internal energy problems broadly stemmed from ignoring the proportionality between temperature and internal energy.

On item 34, in which temperature and internal energy decrease, one student completely ignored the presence of heat transfer out of the system and argued, "…the work done in process 2 [the isobaric compression] would be negative so then the internal energy would ultimately increase." An identical argument was provided by another student who stated, "So for process 2, this will be an isobaric and it looks like negative work is done. So, because work is negative, I'm going to say that it [internal energy] increases." In precisely analogous fashion, when discussing item 45 (an isobaric expansion), another student stated, "I think that [$E$ in isobaric process] would decrease because that is positive work."

It is particularly interesting to contrast students' responses on questions involving adiabatic processes (Section B.4, immediately preceding) and isobaric processes (this section). Many students clearly recognize the absence of heat transfer in an adiabatic process but tend to ignore the effects of work in that context. However, when analyzing isobaric processes—involving horizontal lines on PV diagrams—students seem more easily able to recognize the presence and effects of work, yet more prone to ignore the presence and effects of heat transfer. Based upon these findings, we hypothesize that the term "adiabatic," and the presence of horizontal lines on PV diagrams, respectively, may themselves be key triggers for student thinking that, in many cases, lead students to ignore other important factors that are essential to analyzing the changes that occur during those processes.

6. Incorrectly thinking that $\Delta E=0$ or having incorrect sign of $\Delta E$ for an isochoric process

In isochoric processes, the complete absence of volume changes provides a strong signal to students that work will not be a relevant factor, seemingly forcing them to recognize the presence of heat transfer and the need to analyze its effect. It is possible that due to this, correct-response rates on these items are relatively high and incorrect $\Delta E=0$ responses are relatively infrequent. However, problem context is still a significant influence on student responses, as reflected in responses to items 33, 44, and 68 shown in Table III (Appendix). All three items ask students to determine whether internal energy increases, decreases, or is unchanged in a constant-volume process. Items 33 and 44 can be answered just by examining the PV diagrams which reveal that temperature increases in both cases. Initial and final temperatures are provided explicitly on item 33, but the temperature increase must be inferred from the pressure increase shown in item 44. Predictably, correct-response rates are higher on item 33, but even on item 44, nearly 70% of all introductory students also provide correct responses. However, there is no helpful PV diagram in item 68, and instead students are merely told that "there is net heat transfer to the gas" during a constant volume irreversible process. Still, more than half of introductory students give correct responses on this item. Interviews indicate that incorrect responses are often connected to a failure to recognize the presence or effects of heat transfer. For example, on item 33, one student backtracked by saying, "I feel like internal energy would decrease….Oh, wait, shoot; you're not doing work in process 1, so there would be no change in internal energy." Another student argued along similar lines on item 44: "Well, if there's no work done, there can't be change in internal energy, so there is no change in internal energy in process 1." An interesting mathematical error was reflected in the thinking of an interviewed student who argued, again on item 44, "If I remember correctly, ΔU=PV, oh, sorry, PΔV and in process 1, V is not changing, but P is changing. So if ΔV=0, then P times zero would be an internal energy of zero." The student evidently ignored the contribution of $V\Delta P$ to ΔU.

C. Other difficulties with $W$

Below we describe difficulties other than those associated with $W$ being a path-dependent variable discussed earlier in section III.A.

1. Incorrectly thinking $W \neq 0$ in an irreversible "spontaneous" process

In introductory thermodynamics, the concept of "work" is tied to changes in the configuration of the system in which the system boundary is acted on by an external pressure, i.e., force distributed uniformly around the boundary. In processes in which such a boundary does not exist or cannot move, no work is done. Items 22, 49, and 71 all involve irreversible "spontaneous" processes for which $W = 0$; students are asked to determine whether positive, negative, or zero work is done by the system. Item 22 relates to the free expansion process discussed in an earlier section; gas expands into a vacuum and the absence of force interactions means that no work can be done. Item 49 involves fixed volumes of gases that can exchange thermal energy—heat—but whose volumes can't change so, again, no work is done. Item 71 involves the spontaneous mixing of two containers of gas in which there is no definable system boundary; again, $W = 0$. Results on these items shown in Table IV (Appendix). Almost three-quarters of all introductory students were able to give a correct response on item 49, in which no volume changes of any type occur. However, correct-response rates actually decreased slightly from pre- to post-instruction. Even on item 71 involving the mixing of two gases, about 70% of introductory students gave correct responses; upper-level students did no better.

However, on the free-expansion process, item 22, correct-response rates were much lower: Int-calc, 37%; Int-alg, 39%, and Upper-level, 63%. The most common incorrect response was that positive work would be done by the system (see Ref. [52]), as would have been true had this been a reversible expansion process. Predictably, interview responses showed that students had difficulty in analyzing this "non-standard" expansion process, for example: "I think there's no work done because the gas doesn't act on anything, it's just expanding. But expansion of a gas is also positive work, so that's a little confusing. So I'm just going to say that positive work is done." Other student comments were less ambiguous: "So if the gas is expanding, that means the gas is doing work, so it would be positive"; "Positive work is done by the system, the gas is expanding, it's doing work." Despite evidence suggesting student confusion associated with free-expansion processes (Ref. [36]), there do not appear to be previous reports of detailed investigations of students' thinking regarding *work* in this context. (However, remarks made in passing in Ref. [36] are consistent with our findings.)

Interview responses on item 49 were perhaps less predictable: they suggested that some students thought work was being done despite the explicit problem statement that "the total volume of each chamber remains fixed throughout the process" and that "gas molecules are confined to their original chamber." For example, one student said, "I think the work done by the system would be positive because there needs to be heat going from the higher temperature to the lower temperature, so it would be positive work added to the system." Another student stated, "the work done by the combined system in the two chambers is um…positive since, well actually, heat will flow from hot to cold. So this is taking place freely and it is going…work will be…it will be positive since the gas is moving on its own without an outside source on it." A third student said, "The work done by the combined system of the gasses…ok…they are both expanding into each other. Nothing is being compressed, so the work has to be positive." It appears from these types of interview responses that some students are associating a *movement* by the gas with heat transfer, leading them to think that work is done. It is possible that they are confused by the distinction between the *microscopic* molecular motions that are responsible for heat transfer, and the *macroscopic simultaneous* motion of vast numbers of molecules that are associated with work. Admittedly, this distinction may not often be emphasized in introductory courses.

2. Difficulty in interpreting *W* correctly as area under the curve on a PV diagram

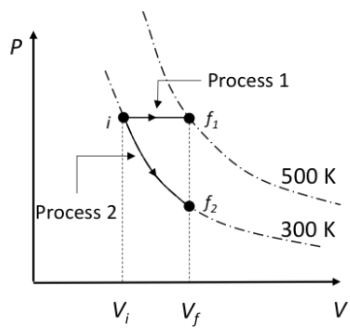
**FIGURE 6.** Diagram provided with items 10 and 11

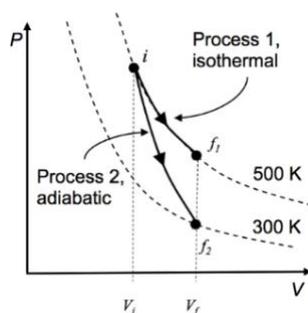
**FIGURE 7.** Diagram provided for items 57-59

One of the most useful problem-solving aids in introductory thermodynamics is an ability to apply the "work done equals area under the curve" interpretation for processes represented on PV diagrams. Student difficulties with this interpretation have been widely reported and analyzed; see, for example, Refs. [22, 23, 56]. Table IV (Appendix) shows results of the survey items that ask students to examine PV diagrams to determine whether work done during a process is positive, negative, or zero. These include items 7 (Fig. 1), 10 (Fig. 6), 42, 43 (Fig. 5 for items 42 and 43), 57, 58 and 59 (Fig. 7 for items 57-59). For all of these, PV diagrams were provided with the problem statement. In each case, the item could be answered merely by a correct application of the "area under the curve" interpretation. Nonetheless, correct-response rates varied widely depending on the specific problem context; they were lowest on item 7, which is the only item that involves a cyclic process (Int-calc: 46%; Int-alg: 18%), and highest on item 43, which involves an isobaric process (74%/60%). (Results for the Int-calc group are consistent with findings reported in Ref. [21].) Item 7 was also the most difficult for the upper-level students (51% correct). One notable difference between the Int-calc and Int-alg students on item 7 is that a sign error was the most common incorrect response given by the Int-calc group, while more than half of the Int-alg students responded that work done would be zero. One of the interview responses on item 7 illustrated a common reasoning process: "If it's one complete cycle, I think it's [work] going to have to be zero because this is the positive work and this is the negative work." The idea that negative work would cancel positive work in a cyclic process was often expressed in the interviews reported in Ref. [23]. This idea is distinct from—although consistent with—the idea discussed in Section A.1 that work done in a cyclic process would be zero since the initial and final volumes are equal.

Items 42, 43, 57, and 58 are similar to each other in that students are asked to determine whether the work done is positive, negative, or zero in non-cyclic processes. The work done in item 42 is zero since it's an isochoric process, while the work is positive in items 43, 57, and 58 which show isobaric, isothermal, and adiabatic processes, respectively. Of the three "positive work" processes, the introductory students clearly found the isobaric-process item easiest to answer. Notably, that process is represented by a horizontal line on the PV diagram while the other two processes are both curved lines, and we may speculate that this horizontal-line feature could have accounted for its higher correct-response rate.

It is interesting and notable that while the Int-calc students found the zero-work (upward-directed vertical line) isochoric-process item no harder than the isobaric-process item, the Int-alg students found it to be much more difficult. Some of the confusion on this item is reflected in written explanations provided by students on their pre-test papers, for example: "An increasing direction means a positive work done by the gas"; "it increases pressure, so it increases work"; "P*V is positive:, P is positive, V is the same"; "The pressure is increasing so the work is positive." Similar thinking was sometimes expressed during the interviews, e.g., one student said, "I would assume that the work done will be positive for process 1. Since the initial here is going up to the pressure to the final point 1."

Items 10 and 59 are similar to each other in that they both involve comparisons of the magnitude of (positive) work done in two different processes that have identical volume changes: isobaric and isothermal in item 10, adiabatic and isothermal in item 59. All three student groups found item 10 (which included the horizontal-line isobaric process) significantly easier to answer than item 59 (in which both processes are represented by curved lines). It may be that the inclusion of the horizontal-line isobaric process is the key feature that makes item 10 the easier one, as also seemed to be the case for item 43. However, confusion regarding adiabatic processes may also have contributed to the greater difficulty of item 59.

Some of the statements made by students during the interviews support the idea that confusion with adiabatic processes may contribute to the relatively low correct-response rate of item 59. For example, one student said, "I think the work would be zero because in adiabatic processes, there's no heat transfer so I don't think it would require work to reach the final state." Written explanations on the pre-test strongly indicated that the relatively larger change in pressure during the adiabatic process (compared to the isothermal process) was a very distracting feature that helped to prompt sign errors, for example: "The adiabatic process experiences a larger change in P and volume"; "The pressure of the adiabatic system is less than that isothermal system"; "There is a greater difference in pressure loss, the process with a greater pressure loss experienced more work done." Since the PV diagram makes it evident that there is more "area under the curve" in the isothermal process, it is clear that the area interpretation was often inadequate to counter the chains of "intuitive" reasoning triggered by other features of the diagram.

Although only about 10% of students thought that work done in the two processes in item 10 (see Fig. 6) would be equal, an idea expressed during two of the interviews is worth noting, e.g., one student reasoned, "I'm thinking work equals P$\Delta$V. They both start at the same P and then both end at the same $V_f$, so I'm thinking that the work done would be equal." Another student stated, "…the start volume is the same and the end volume is the same, so the work done is the same."

3. Difficulties with evaluating work in the absence of a PV diagram

At the introductory level, students are expected to learn that work done by the system is positive during an expansion and negative during a compression. Apart from that, comparisons of work done in different processes would normally be done by reference to a PV diagram, although of course one could also make use of analytic expressions for work in terms of P and V. Items 3 and 62 lack PV diagrams and simply ask whether the work done by a gas would be positive, negative, or zero for a "reversible adiabatic expansion process" (item 3) and "an isothermal expansion" (item 62). Although one might have expected the word "expansion" to immediately trigger the response "positive work done," the post-instruction correct-response rate on these items (see Table IV, Appendix) was not particularly high for introductory students, ranging from 41% for Int-alg on item 3 to 65% for Int-calc on item 62. (By comparison, at least three-quarters of upper-level students answered items correctly on both items 3 and 62.)

It is difficult to say whether the absence of PV diagrams was a factor, since correct-response rates were comparable for items that *were* accompanied by PV diagrams. It is notable that more than a quarter of the Int-alg students answered that zero work would be done during the adiabatic expansion (item 3). (Some interviews indicated a confusion between "zero heat" and "zero work" for adiabatic processes.) Zero-work responses were less frequent on item 62, but sign errors were made on both items; these sign errors (stating work done would be negative for the expansions) were made by about a quarter of introductory students on these two items. Some of the explanations offered by students who provided written explanations on the pre-test were reminiscent of those reported in Ref. [23], for example: "work is done to the gas for it to expand"; "the gas expands so is doing negative work"; "the work done by the heat is positive, so the work done by the gas is negative." These responses illustrate the potential importance of sign confusion in work problems.

D. Other difficulties with $Q$

In contrast to work, there is no direct way to infer the sign of heat transfer merely by examining superficial features of a PV diagram. There are special cases that one might think should be easier, in particular, adiabatic processes (in which heat transfer is by definition zero) and isochoric processes (in which no work is done, so that heat transfer will be positive or negative depending on the sign of the temperature/pressure change). Other than that, first-law considerations involving $\Delta E = Q - W$ must be invoked. Below we describe survey results on several items that reflect these varied conditions. (Also see previous discussion in section III.A.)

1. $Q = 0$ or incorrect sign of $Q$ for an isothermal process

One of the more remarkable and robust findings of several previous investigations is a widespread student alternative conception that there is no heat transfer during an isothermal process. In this sense, students are confusing the distinction between isothermal and adiabatic processes; see, for example, Refs. [22, 28, 30, 39, 47]. Consistent with these prior studies, we found that many students could not clearly distinguish in practice between an isothermal and an adiabatic process, thus being led to the conclusion that there is no heat transfer in an isothermal process. On items 4 and 32 (the latter a true/false question), both describing isothermal processes, about half of all introductory students responded that net heat transfer would be zero. On item 60, involving an "isothermal expansion," about one third of introductory students gave the $Q = 0$ response. On item 4, there was—unexpectedly—a marked decline in the performance of the Int-calc students from pre- to post-instruction, with the prevalence of the $Q = 0$ response increasing sharply. It is possible that the presence of the modifier "reversible" in item 4 (describing a "reversible isothermal compression") became more of a distractor for the Int-calc students after instruction in relevant concepts in their introductory college physics courses, a cautionary lesson regarding the possible effect of poorly understood terminology infiltrating students' thinking. On item 60, there was no overall performance decline after instruction, but the wide prevalence of the $Q = 0$ idea changed only slightly. Sign errors on items 4 and 60 remained in the 15-20% range both before and after instruction. (These included responses of $Q > 0$ during an isothermal compression and $Q < 0$ during an expansion, the opposite of what actually occurs.)

Comments from student interviews clearly illustrated the thinking behind the incorrect responses, and were remarkably similar to those reported in previous investigations such as Ref. [23]; e.g., here are illuminating comments on item 4: "Isothermal is no change in temperature, so I think there is no net heat transfer because Q = mc $\Delta$T and $\Delta$T is zero"; "I believe there's not a heat transfer since isothermal process is the one that has the constant temperature so there's no net heat transfer to the gas"; "There shouldn't be any heat transfer because it is isothermal"; "So if it's isothermal, the temperature stays constant which would mean there would be no heat transfer." Similarly, on item 60, a student stated, "So it's isothermal. I feel like there is no heat transfer because you are not changing the temperature…Isothermal yeah, yeah, I don't think there is heat transfer." Similar arguments were advanced for item 32: "For isothermal, that is the same temperature. The temperature is constant, so I don't think there would be a heat transfer." Regarding sign errors, interview responses suggest that many students may have failed to invoke the first law of thermodynamics and/or ignored the role of work by the gas.

2. $Q \neq 0$ in an adiabatic process

The survey explicitly included the meaning of the word adiabatic on the cover page with the other instructions. Since $Q = 0$ by definition in an adiabatic process, one might expect item 31 (true/false: is there no net heat transfer in an adiabatic process) to be the easiest of this group, and indeed it does show the highest post-instruction correct-response rate of 69% for all introductory students and 87% for upper level. Still, these figures reflect a large number of incorrect student responses. Students were even less successful on item 39, which involves a gas being heated by a burner; students are asked whether or not this is an adiabatic process. Only 61%/48% of the introductory (calc/alg) students gave correct "no" responses. Some introductory students who provided written explanations on the pre-test for item 31 provided explanations such as, "Heat can be transferred out of an adiabatic process"; "A process being adiabatic does not prevent net heat transfer out of the system"; and "Adiabatic involves the cooling down of a system." As for explanations of why a heating process could be described as adiabatic, one interviewee said, "Adiabatic, there is no heat transfer. That is also possible since no thermal contact with anything else." It is possible that this student might have thought that thermal contact requires conduction through direct contact between solid objects, a misunderstanding that merits further investigation. Other explanations suggested that some students had simply forgotten or ignored the basic definition of the term "adiabatic." Our findings are consistent with those of prior studies that identified difficulties with adiabatic processes, such as Refs. [22, 26, 27, 30]. However, our explicit inclusion of the *definition* of adiabatic on the survey instrument and the unambiguous interview responses we obtained suggest that for many students, the term "adiabatic" conveys no (or negative) useful information. Implications for instruction are discussed below.

3. Difficulty with $Q$ for an isobaric expansion

Item 41 (Fig. 5) asks students to decide whether net heat transfer in an isobaric expansion (represented by a horizontal line on a PV diagram) was to the gas, away from the gas, or zero. Solution of this problem requires use of the first law in a manner similar to the isothermal process items, without the evident distraction of being a "constant temperature" process. Nonetheless, the low correct-response rates were quite comparable to those on the isothermal process item 60. However, sign errors were more prevalent on item 41 and $Q = 0$ errors much less so, in comparison to item 60. Interview responses on this item were not very enlightening and incorrect responses reflected general confusion.

A related "Yes/No" item is 38, which describes an isobaric expansion without identifying the process as such. A diagram is shown of a gas in a closed container being heated with a burner, such that the top of the container is a slowly moving piston "that can move freely, without friction" and thus the gas is expanding; there is no other thermal contact between the gas and the environment. Students are asked whether the process can be described as isothermal. (Items 37 and 39 ask whether the description could be "constant pressure (isobaric)" or "adiabatic," respectively.) Students are expected to recognize that the process is an isobaric expansion with $PV$ increasing, thus implying a temperature increase and excluding the "isothermal" description. Correct-response rates are around 70%, but around 30% of students thus accepted the isothermal label. Interviews revealed "compensatory" thinking in which work done is believed to balance the heat transferred; for example, one student said, "possible [that it could be isothermal] if the…yeah if the work done compensates the Q, that could happen." However, this type of reasoning ignores the ideal gas law $PV = NkT$ that is printed right at the beginning of the survey for students' reference.

4. Difficulty with $Q$ for an isochoric process

Item 40 shows an isochoric process on a PV diagram and asks whether heat transfer is to the gas, away from the gas, or zero. Pressure is shown to be increasing, implying that the temperature also is rising and therefore that there must be heat transfer to the gas. Correct response rates were 61%/58% for the introductory students (calc/alg), and 83% for upper-level students. Incorrect responses for the Int-calc students were split evenly between sign errors (stating heat transfer was *away* from the gas) and claims that $Q = 0$. Among Int-alg, sign errors were more common. Explanations provided on written pretests indicated that many students simply ignored work and the first law and relied on various "intuitive" ideas. Prior research investigated student idea that $Q = 0$ in an isochoric process [28, 30, 50, 51], and less focus on the sign errors that we find and describe here.

## IV. SUMMARY AND INSTRUCTIONAL IMPLICATIONS

Using a validated survey instrument administered both pre- and post-instruction, we studied the difficulties encountered by both introductory and upper-level students with concepts related to changes in internal energy, work done by and on a system, and heat transfer to and from a system, at the level and depth typical of introductory college physics courses. The post-instruction administration came after traditional lecture-based instruction. The findings presented here suggest that even though instructors may cover all relevant topics in class, concepts related to internal energy, work, and heat transfer can continue to be challenging even for upper-level students.

Our findings reflect responses from more than a thousand students enrolled in introductory and upper-level courses (12 different courses) from four different universities, with test items spanning 19 different problem contexts. They are consistent with previously reported findings by studies in similar instructional settings, using similar problem contexts. In particular, some of our survey items are similar to or the same as those used in previous investigations, while others are new. Thus, our research validates the previous findings in new problem contexts and points to the robustness of the previous findings. Our findings expand on those of previous investigations even in identical problem contexts because we have focused on introductory algebra-based, introductory calculus-based, and advanced-level students, all in the same study, a unique feature of the current work. Thus, our findings can also be useful for gaining insight into how certain previous findings may generalize across different levels of students. Moreover, our findings reflecting the context dependence of student reasoning on these thermodynamics concepts can also be valuable for thermodynamics educators who are planning curriculum and instruction.

For reference, we enumerate here a few of our findings that have not been—to our knowledge—previously reported in the physics education literature: (1) prevalence and nature of student difficulties related to heat transfer and change in internal energy in cyclic processes as represented on PV diagrams; (2) extension of previously reported findings on difficulties with work in cyclic processes to diverse types of PV diagrams; (3) the difficulty with the $E$-$T$ proportionality in ideal gases is manifested by a strong context dependence of correct-response rates, even when temperatures are *explicitly* provided; (4) evidence that many students believe that work is done in free-expansion processes (providing strong confirmation of inconclusive observations reported in some previous studies) and affirmation that this idea also leads students astray when considering changes in internal energy; (5) ignoring heat transfer in isothermal processes (a previously reported finding) can lead students to conclude both that internal energy in such processes *decreases* (due to work being done) or *increases* (due to an association between increasing volume and greater internal energy); (6) idea that in an adiabatic expansion, internal energy might be *constant* (misidentifying "adiabatic" with "constant energy"); (7) many students clearly recognize the absence of heat transfer in an adiabatic process but tend to ignore the effects of work in that context; however, when analyzing isobaric processes—involving horizontal lines on PV diagrams—students seem more easily able to recognize the presence and effects of work, yet more prone to ignore the presence and effects of heat transfer; (8) students often fail to recognize the presence or effects of heat transfer in isochoric processes; (9) students were more successful in finding the sign of work done in isobaric than in other processes, but tended to associate increasing pressure with positive work in isochoric processes; (10) students frequently asserted that heat transfer could or would occur during processes explicitly identified as adiabatic.

Instructors can view the wide variety of thermodynamics problem contexts utilized in the survey as a means for helping students develop a functional understanding of these concepts, at the same time helping them develop problem-solving and reasoning skills. Instructors should consider using physics education research based instructional approaches to improve the effectiveness of curricula and pedagogical methods in helping students learn key concepts [28, 56]. The findings presented here regarding student difficulties in traditional lecture-based introductory and upper-level courses can be used as baseline data, to be compared with courses in which innovative evidence-based curricula and pedagogies are used, to gauge the level of improvement in introductory and advanced students' understanding of these concepts.

## ACKNOWLEDGMENTS

We thank the students and faculty from all of the universities who helped with this research.

## APPENDIX A: TABLES FOR PRE- AND POST-TEST SCORES

TABLE II. Correct-response rates as percentage of total responses (in boldface) and rates of various incorrect responses for students in upper-level, introductory calculus-based (Int-calc), and algebra-based (Int-alg) physics courses for problems related to changes in internal energy. The survey item numbers and choices that constitute a particular row in the table are provided. Item #s with an asterisk (*) are T/F questions.

| **Correct answers in bold**, incorrect answers unbolded | Item # | Answer Choices | Prevalence (%) | | | | |
|---|---|---|---|---|---|---|---|
| | | | Upper Post | Int-calc Post | Int-calc Pre | Int-alg Post | Int-alg Pre |
| **Correct responses to ΔE questions** | **6** | **A** | **89** | **80** | **79** | **80** | **80** |
| | **13** | **C** | **94** | **82** | **84** | **83** | **79** |
| | **23** | **A** | **76** | **57** | **60** | **58** | **55** |
| | **25** | **C** | **83** | **72** | **63** | **69** | **72** |
| | **33** | **B** | **83** | **76** | **84** | **81** | **85** |
| | **34** | **C** | **70** | **58** | **59** | **55** | **58** |
| | **47*** | **A** | **84** | **71** | **72** | **72** | **73** |
| | **50** | **C** | **89** | **69** | **78** | **77** | **78** |
| | **69** | **A** | **60** | **32** | **34** | **25** | **27** |
| | **72** | **C** | **78** | **60** | **72** | **63** | **66** |
| Not treating the internal energy E as a state variable | 6 | B+C+D | 11 | 20 | 21 | 20 | 20 |
| | 25 | A+B+D | 17 | 28 | 38 | 31 | 28 |
| | 47* | B | 16 | 29 | 28 | 28 | 27 |
| Not recognizing that E is proportional to T for an ideal gas | 33 | A+C+D | 17 | 24 | 16 | 19 | 15 |
| | 34 | A+B+D | 30 | 42 | 41 | 45 | 42 |
| | 69 | B+C+D | 40 | 68 | 66 | 75 | 73 |
| ΔE≠0 for an isolated system | 13 | A+B+D | 6 | 18 | 16 | 17 | 21 |
| | 23 | B+C+D | 24 | 43 | 40 | 42 | 45 |
| | 50 | A+B+D | 11 | 31 | 22 | 23 | 22 |
| | 72 | A+B+D | 22 | 40 | 28 | 37 | 34 |
| ΔE>0 for an isolated system | 13 | A | 3 | 6 | 7 | 8 | 12 |
| | 23 | B | 6 | 14 | 10 | 11 | 13 |
| | 50 | A | 8 | 17 | 7 | 7 | 7 |
| | 72 | A | 7 | 21 | 15 | 17 | 15 |
| ΔE<0 for an isolated system | 13 | B | 2 | 9 | 6 | 8 | 6 |
| | 23 | C | 15 | 28 | 29 | 27 | 31 |
| | 50 | B | 3 | 13 | 12 | 14 | 11 |
| | 72 | B | 13 | 17 | 11 | 17 | 17 |

TABLE III. Correct-response rates as percentage of total responses (in boldface) and rates of various incorrect responses for students in upper-level, introductory calculus-based (Int-calc), and algebra-based (Int-alg) physics courses for problems related to changes in internal energy. The survey item numbers and choices that constitute a particular row in the table are provided.

| **Correct answers in bold**, incorrect responses unbolded | Item # | Answer Choices | Prevalence (%) | | | | |
|---|---|---|---|---|---|---|---|
| | | | Upper Post | Int-calc Post | Int-calc Pre | Int-alg Post | Int-alg Pre |
| **Correct responses to ΔE questions** | **2** | **A** | **54** | **30** | **20** | **24** | **28** |
| | **33** | **B** | **83** | **76** | **84** | **81** | **85** |
| | **34** | **C** | **70** | **58** | **59** | **55** | **58** |
| | **44** | **B** | **88** | **69** | **69** | **70** | **74** |
| | **45** | **B** | **65** | **43** | **32** | **37** | **32** |

| | Item # | Answer Choices | Upper Post | Int-calc Post | Int-calc Pre | Int-alg Post | Int-alg Pre |
|---|---|---|---|---|---|---|---|
| | **61** | **A** | **64** | **48** | **34** | **28** | **27** |
| | **65** | **A** | **69** | **49** | **34** | **40** | **36** |
| | **68** | **B** | **70** | **55** | **55** | **61** | **58** |
| ΔE≠0 for an isothermal expansion (ideal monatomic gas) | 61 | B+C+D | 36 | 52 | 66 | 72 | 73 |
| ΔE>0 for an isothermal expansion (ideal monatomic gas) | | B | 24 | 26 | 36 | 39 | 44 |
| ΔE<0 for an isothermal expansion (ideal monatomic gas) | | C | 7 | 22 | 26 | 28 | 26 |
| ΔE≠0 for an isothermal expansion (ideal monatomic gas) | 65 | B+C+D | 31 | 51 | 66 | 60 | 64 |
| ΔE>0 for an isothermal expansion (ideal monatomic gas) | | B | 22 | 27 | 32 | 35 | 38 |
| ΔE<0 for an isothermal expansion (ideal monatomic gas) | | C | 6 | 23 | 30 | 21 | 25 |
| ΔE=0 for reversible adiabatic expansion | 2 | C | 20 | 39 | 46 | 47 | 43 |
| ΔE>0 for reversible adiabatic expansion | | B | 21 | 28 | 32 | 26 | 28 |
| ΔE=0 for an isobaric compression (ideal monatomic gas) | 34 | A | 7 | 12 | 24 | 20 | 24 |
| ΔE>0 for an isobaric compression (ideal monatomic gas) | | B | 15 | 25 | 15 | 20 | 17 |
| ΔE=0 for an isobaric expansion (ideal monatomic gas) | 45 | A | 7 | 17 | 29 | 22 | 29 |
| ΔE<0 for an isobaric expansion (ideal monatomic gas) | | C | 20 | 33 | 34 | 37 | 36 |
| ΔE=0 for an isochoric process with $P_f > P_i$ (ideal monatomic gas) | 33 | A | 8 | 16 | 9 | 11 | 8 |
| ΔE<0 for an isochoric process with $P_f > P_i$ (ideal monatomic gas) | | C | 7 | 8 | 6 | 6 | 6 |
| ΔE=0 for an isochoric process with $P_f > P_i$ (ideal monatomic gas) | 44 | A | 7 | 18 | 15 | 17 | 14 |
| ΔE<0 for an isochoric process with $P_f > P_i$ (ideal monatomic gas) | | C | 4 | 8 | 12 | 11 | 8 |
| ΔE=0 for an isochoric process with Q>0 (ideal monatomic gas) | 68 | A | 14 | 24 | 17 | 18 | 18 |
| ΔE<0 for an isochoric process with Q>0 (ideal monatomic gas) | | C | 9 | 17 | 23 | 17 | 19 |

Table IV. Correct-response rates as percentage of total responses (in boldface) and rates of various incorrect responses for students in upper-level, introductory calculus-based (Int-calc), and algebra-based (Int-alg) physics courses for problems related to work. The survey item numbers and choices that constitute a particular row in the table are provided. Item #s with an asterisk (*) are T/F questions.

| | | | Prevalence (%) | | | | |
|---|---|---|---|---|---|---|---|
| **Correct answers in bold**, incorrect answers unbolded | Item # | Answer Choices | Upper Post | Int-calc Post | Int-calc Pre | Int-alg Post | Int-alg Pre |
| | **3** | **A** | **75** | **54** | **48** | **41** | **41** |
| | **7** | **B** | **51** | **46** | **13** | **18** | **5** |
| | **10** | **B** | **88** | **72** | **55** | **58** | **54** |
| | **22** | **A** | **63** | **37** | **39** | **39** | **36** |
| | **42** | **C** | **84** | **74** | **28** | **46** | **17** |
| | **43** | **A** | **83** | **74** | **45** | **60** | **43** |
| **Correct responses to W questions** | **46*** | **B** | **90** | **74** | **63** | **68** | **65** |
| | **49** | **A** | **89** | **72** | **79** | **75** | **78** |
| | **57** | **B** | **89** | **64** | **28** | **52** | **27** |
| | **58** | **B** | **87** | **61** | **32** | **44** | **23** |
| | **59** | **B** | **78** | **54** | **29** | **38** | **30** |
| | **62** | **A** | **81** | **65** | **46** | **55** | **47** |
| | **71** | **A** | **71** | **67** | **75** | **70** | **77** |
| W = 0 in a complete counterclockwise cycle (PV diagram) | 7 | C | 16 | 21 | 59 | 51 | 71 |
| W > 0 in a complete counterclockwise cycle (PV diagram) | | A | 31 | 31 | 26 | 28 | 22 |
| W is determined by the state of the system | 46* | A | 10 | 26 | 37 | 32 | 35 |
| W ≠ 0 for a spontaneous process with only particles transfer | 22 | B+C+D | 38 | 63 | 61 | 61 | 64 |
| | 71 | B+C+D | 29 | 33 | 25 | 30 | 23 |
| W ≠ 0 for a spontaneous process with only heat transfer | 49 | B+C+D | 11 | 28 | 21 | 25 | 22 |
| | 7 | A+C+D | 49 | 54 | 87 | 82 | 95 |

| Description | Item # | Answer Choices | Upper Post | Int-calc Post | Int-calc Pre | Int-alg Post | Int-alg Pre |
|---|---|---|---|---|---|---|---|
| Not deciphering W as area under the curve on a PV diagram correctly | 10 | A+C+D | 12 | 28 | 45 | 42 | 46 |
| | 42 | A+B+D | 16 | 26 | 72 | 54 | 83 |
| | 43 | B+C+D | 17 | 26 | 55 | 40 | 57 |
| | 57 | A+C+D | 11 | 36 | 72 | 48 | 73 |
| | 58 | A+C+D | 13 | 39 | 68 | 56 | 77 |
| | 59 | A+C+D | 22 | 46 | 71 | 62 | 70 |
| W = 0 in an isothermal expansion (PV diagram) | 57 | A | 3 | 10 | 24 | 15 | 21 |
| W < 0 in an isothermal expansion (PV diagram) | | C | 7 | 24 | 43 | 33 | 49 |
| W = 0 in an isothermal expansion | 62 | C | 4 | 11 | 22 | 16 | 21 |
| W < 0 in an isothermal expansion | | B | 11 | 19 | 28 | 24 | 26 |
| W = 0 in a reversible adiabatic expansion | 3 | C | 11 | 19 | 27 | 27 | 28 |
| W < 0 in a reversible adiabatic expansion | | B | 10 | 24 | 19 | 27 | 25 |
| W = 0 in an adiabatic expansion (PV diagram) | 58 | A | 6 | 10 | 10 | 13 | 17 |
| W < 0 in an adiabatic expansion (PV diagram) | | C | 7 | 27 | 54 | 41 | 59 |
| W = 0 in an isobaric expansion (PV diagram) | 43 | C | 4 | 9 | 26 | 17 | 28 |
| W < 0 in an isobaric expansion (PV diagram) | | B | 12 | 16 | 26 | 21 | 27 |
| *Comparing isothermal expansion to adiabatic expansion when both start in same state and have same ΔV (PV diagram)* | | | | | | | |
| $W_{adi} = W_{isthm}$ | 59 | A | 9 | 17 | 14 | 14 | 11 |
| $W_{adi} > W_{isthm}$ | | C | 12 | 26 | 51 | 44 | 53 |

Table V. Correct-response rates as percentage of total responses (in boldface) and rates of various incorrect responses for students in upper-level, introductory calculus-based (Int-calc), and algebra-based (Int-alg) physics courses for problems related to heat transfer. The survey item numbers and choices that constitute a particular row in the table are provided. Item #s with an asterisk (*) are T/F questions.

| **Correct answers in bold**, difficulties unbolded | Item # | Answer Choices | Prevalence (%) | | | | |
|---|---|---|---|---|---|---|---|
| | | | Upper Post | Int-calc Post | Int-calc Pre | Int-alg Post | Int-alg Pre |
| **Correct responses to Q questions** | **4** | **B** | **68** | **30** | **42** | **21** | **28** |
| | **9** | **C** | **44** | **35** | **21** | **20** | **16** |
| | **12** | **C** | **47** | **29** | **23** | **23** | **17** |
| | **26** | **B** | **52** | **31** | **16** | **16** | **18** |
| | **30*** | **B** | **85** | **67** | **55** | **55** | **50** |
| | **31*** | **A** | **87** | **69** | **51** | **69** | **53** |
| | **32*** | **B** | **84** | **54** | **45** | **50** | **40** |
| | **39** | **B** | **88** | **61** | **52** | **48** | **44** |
| | **40** | **B** | **83** | **61** | **55** | **58** | **60** |
| | **41** | **B** | **67** | **45** | **37** | **42** | **43** |
| | **60** | **B** | **75** | **43** | **34** | **44** | **35** |
| Q = 0 in a complete counterclockwise cycle (PV diagram) | 9 | A | 16 | 33 | 44 | 56 | 58 |
| Q > 0 in a complete counterclockwise cycle (PV diagram) | | B | 29 | 26 | 22 | 12 | 16 |
| Q = 0 in a complete clockwise cycle (PV diagram) | 26 | A | 25 | 37 | 49 | 55 | 61 |
| Q < 0 in a complete clockwise cycle (PV diagram) | | C | 13 | 28 | 24 | 24 | 17 |
| Q = 0 in any complete cycle | 30* | A | 15 | 33 | 45 | 45 | 50 |
| Treating Q as if it were a state variable (PV diagram) | 12 | A | 28 | 56 | 67 | 66 | 72 |
| Q = 0 in reversible isothermal compression | 4 | C | 17 | 50 | 30 | 59 | 51 |
| Q > 0 in reversible isothermal compression | | A | 13 | 18 | 27 | 18 | 19 |
| Q = 0 in an isothermal expansion | 60 | A | 12 | 30 | 39 | 37 | 37 |
| Q < 0 in an isothermal expansion | | C | 10 | 20 | 24 | 15 | 24 |
| Q = 0 in any isothermal process | 32* | A | 16 | 46 | 55 | 50 | 60 |

| | | | | | | |
|---|---|---|---|---|---|---|
| Q ≠ 0 in an adiabatic process | 31* | B | 13 | 31 | 49 | 31 | 47 |
| Q = 0 in an isobaric expansion (PV diagram) | 41 | A | 9 | 17 | 25 | 20 | 24 |
| Q < 0 in any isobaric expansion (PV diagram) | | C | 17 | 32 | 31 | 31 | 26 |
| Q = 0 in an isochoric process with $P_f > P_i$ (PV diagram) | 40 | A | 9 | 18 | 11 | 13 | 14 |
| Q < 0 in an isochoric process with $P_f > P_i$ (PV diagram) | | C | 3 | 18 | 27 | 24 | 20 |
| A gas heated in a closed container with a slowly moving piston with no other thermal contact can be an isothermal process | 38* | A | 22 | 28 | 53 | 33 | 45 |
| A gas heated in a closed container with a slowly moving piston with no other thermal contact can be an adiabatic process | 39* | A | 13 | 39 | 48 | 52 | 56 |